\documentclass[a4paper,twoside,12pt]{article}
\input{obsjkt.sty}

\usepackage{rotating}

\begin{document} 

\OBSheader{Lagrangian point positions}{J.\ Southworth}{2023 August}

\OBStitle{Simple approximations to the positions of the Lagrangian points}

\OBSauth{John Southworth}

\OBSinstone{Astrophysics Group, Keele University, Staffordshire, ST5 5BG, UK}

\OBSabstract{The Roche potential is the sum of the gravitational and rotational potentials experienced by a massless body rotating alongside two massive bodies in a circular orbit. The Lagrangian points are five stationary points in the Roche potential. The positions of two of the Lagrangian points (L4 and L5) are fixed. The other three (L1, L2 and L3) are along the line joining the two masses: their positions depend on the mass ratio, $q$, and can be calculated numerically by finding the roots of a quintic polynomial. Analytical approximations to their positions are useful in several situations, but existing ones are designed for small mass ratios. We present new approximations valid for all mass ratios from zero to unity:
\begin{eqnarray*}
x_{\rm L1} & = & 1 - \frac{q^{0.33071}}{0.51233\,q^{0.49128} + 1.487864} \\
x_{\rm L2} & = & 1 + \frac{q^{0.8383} +  2.891\,q^{0.3358}}{1.525\,q^{0.848} + 4.046596} \\
x_{\rm L3} & = & -1 + \frac{q^{1.007}}{1.653\,q^{0.9375} + 1.66308}
\end{eqnarray*}
in a rotating frame of reference where the more massive body is at $x=0$ and the less massive body at $x=1$. The three approximations are precise to $6 \times 10^{-5}$ for all mass ratios.
}


\section*{The Roche potential}

The motion of two point masses in orbit around each other is a well-known and well-understood\cite{Kopal59book,Bate++71book,Hilditch01book,Roy05book} solved problem in celestial mechanics. Its extension to three bodies was a major goal of 19th-century astrophysics, until Henri Poincar\'e demonstrated in 1887 that it was insoluble.

There is one situation where analytically tractable results can be obtained, known as the \emph{restricted three-body problem}. In this case, termed the \emph{Roche model}, two point masses are on a circular orbit and the third body is massless. The Roche potential is the sum of the gravitational and rotational potentials and can be used to describe the shapes of stars in close binary systems. The more massive star is defined to be at the origin of a Cartesian co-ordinate system ($x=y=z=0$) and the less massive star is at $(x,y,z)=(1,0,0)$. The co-ordinate system therefore rotates with the binary system.

The Roche potential experienced by a massless particle at point ($x$,$y$,$z$) is
\begin{equation}
\Phi = - \frac{GM_1}{r_1} - \frac{GM_2}{r_2} - \frac{G(M_1+M_2)}{2}\left[\left(x-\frac{M_2}{M_1+M_2}\right)^2+y^2\right]
\end{equation}
where $G$ is the Newtonian gravitational constant, $M_1$ is the mass of the more massive star, $M_2$ is the mass of the less massive star, $r_1$ is the distance from the point to mass $M_1$, and $r_2$ is the distance from the point to mass $M_2$. Simple trigonometry gives
\begin{equation}
r_1 = \sqrt{x^2+y^2+z^2}
\end{equation}
and
\begin{equation}
r_2 = \sqrt{(1-x)^2+y^2+z^2}
\end{equation}

It is convenient to work with the normalised Roche potential
\begin{equation}
\Phi_{\rm n} = \frac{-2}{G(M_1+M_2)}\Phi
\end{equation}
which yields
\begin{equation} \label{eq:phin}
\Phi_{\rm n} = \frac{2}{r_1}\frac{1}{1+q} + \frac{2}{r_2}\frac{q}{1+q} + \left(x-\frac{q}{1+q}\right)^2 + y^2
\end{equation}
where $q = M_2 / M_1$ is the mass ratio. Thus the Roche potential has been simplified into a function which depends only on the position of the massless body, the mass ratio, and the scale of the system which is set by the semimajor axis of the relative orbit of the two massive bodies.

\begin{figure}[t] \centering \includegraphics[width=\textwidth]{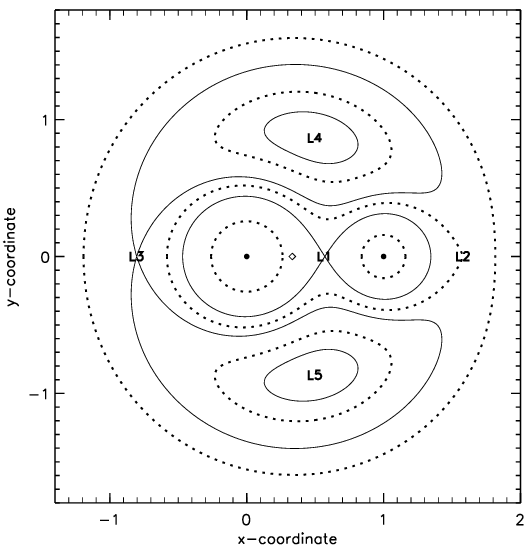} \\
\caption{\label{fig:roche} Visualisation of the Roche potential for a system with
$q=0.5$. Lines of constant potential are shown with solid or dotted lines. The
positions of the two masses are shown with filled circles. The position of the centre
of mass is shown with an open diamond. The Lagrangian points are labelled.} \end{figure}


\section*{The Lagrangian points}

There are five stationary points in the Roche potential where no net force is exerted on a particle, i.e.\ the gravitational and rotational forces balance. These are called the Lagrangian points and will be denoted L1, L2, L3, L4 and L5 below. The first three were discovered by Leohard Euler and the last two by Joseph-Louis Lagrange. All five Lagrangian points are in the $x$--$y$ plane so have $z=0$. Fig.~\ref{fig:roche} shows their positions in the case that $q=0.5$.

The positions of the L4 and L5 points each form an equilateral triangle with the two masses at the other vertices. Their positions are thus fixed at
\begin{equation}
x_{\rm L4} = \frac{1}{2} \qquad \qquad y_{\rm L4} = \frac{\sqrt{3}}{~~2}
\end{equation}
\begin{equation}
x_{\rm L5} = \frac{1}{2} \qquad \qquad y_{\rm L5} = \frac{-\sqrt{3}}{~~2}
\end{equation}
and need not be discussed further. Orbits at the L4 and L5 points are stable for mass ratios $q < 0.04004$ (Pr\v{s}a \cite{Prsa18book}).

\begin{figure}[t] \centering \includegraphics[width=\textwidth]{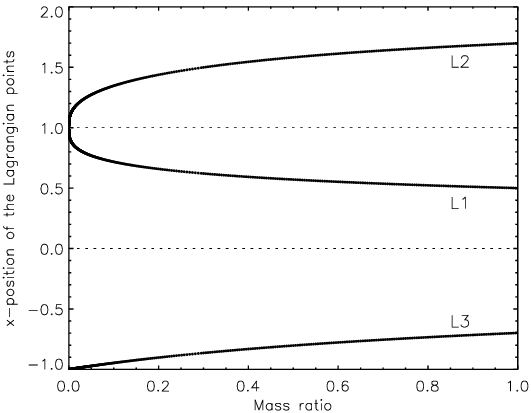} \\
\caption{\label{fig:l1l2l3} $x$-positions of the L1, L2 and L3 points as a function of mass ratio.} \end{figure}

The L1 (inner Lagrangian), L2 (outer Lagrangian) and L3 points are all along the $x$-axis (so $y=z=0$) but their positions depend on mass ratio and are less straightforward to determine. By setting the derivative of Eq.~\ref{eq:phin} to zero we find
\begin{equation}
0 = -\frac{1}{x^2} + \frac{q}{(1-x)^2} + (1+q)x - q
\end{equation}
where $r_1 \to x$ and $r_2 \to 1-x$. Further algebra leads to the equation
\begin{equation} \label{eq:quintic}
0 = -1 + 2x - x^2 + (1+3q)x^3 - (2+3q)x^4 + (1+q)x^5
\end{equation}
which is quintic so does not have a general solution. Eq.~\ref{eq:quintic} can be solved using numerical methods, e.g.\ bisection, Newton-Raphson or grid search (see Leahy \& Leahy \cite{LeahyLeahy15comac}). The positions of the L1, L2 and L3 points are shown as a function of mass ratio in Fig.~\ref{fig:l1l2l3}.

\section*{Existing approximations to the positions of the L1, L2 and L3 points}

There are times when a precise solution of Eq.~\ref{eq:quintic} is not necessary, and a quick and simple approximation is adequate. These cases include when making a plot, trying to put together the first version of a piece of code, or as a starting estimate for iterative refinement. Approximations are available\footnote{https://en.wikipedia.org/wiki/Lagrange\_\,point}\footnote{https://map.gsfc.nasa.gov/ContentMedia/lagrange.pdf}, but in general are only valid for small mass ratios.

At small mass ratios the distances of the L1 and L2 points from the second mass become approximately equal to the size of its Hill sphere (e.g.\ Ref.~\cite{DepaterLissauer01book}):
\begin{equation}
x_{\rm L1} = 1 - \left(\frac{q}{3}\right)^{1/3}
\qquad \qquad
x_{\rm L2} = 1 +  \left(\frac{q}{3}\right)^{1/3}
\end{equation}

Another set of approximations was reported in the textbook by Pr\v{s}a \cite{Prsa18book}, and come from a perturbation analysis presented in the textbook by Battin \cite{Battin87book}. In the geometry adopted in the current work these are:
\begin{equation} \label{eq:taff:l1}
x_{\rm L1} = 1 - \zeta + \frac{1}{3}\zeta^2 + \frac{1}{9}\zeta^3 - \frac{58}{81}\zeta^4
\end{equation}
\begin{equation} \label{eq:taff:l2}
x_{\rm L2} = 1 + \zeta + \frac{1}{3}\zeta^2 - \frac{1}{9}\zeta^3 + \frac{58}{81}\zeta^4
\end{equation}
\begin{equation} \label{eq:taff:l3}
x_{\rm L3} = -1 + \frac{7}{12}\mu + \frac{1127}{20736}\mu^3 + \frac{7889}{248832}\mu^4
\end{equation}
where $\mu = q/(1+q)$ and $\zeta = (\mu/3)^{1/3}$. These are designed for small mass ratios so are imprecise for larger values. The deviations for the positions of L1, L2 and L3 are less than $10^{-5}$ for $q$ $<$ 0.01, 0.03 and 0.26, respectively. The largest deviations are $-0.035$ at $q=1$, $-0.0014$ at $q=0.41$, and 0.0012 at $q=1$, respectively. The largest deviations can be decreased by optimising the numerical coefficients in eqs.\ \ref{eq:taff:l1} to \ref{eq:taff:l3} at the expense of lowering the quality of the approximations for small mass ratio.

Similar approximations to those in Eqs.\ \ref{eq:taff:l1} to \ref{eq:taff:l3} were given in Murray \& Dermott \cite{MurrayDermott00book} (pp.\ 78-80). They work well for small mass ratios, but deviate at higher mass ratios by much more than do Eqs.\ \ref{eq:taff:l1} to \ref{eq:taff:l3}, so will not be considered further. Other approximations suitable for small mass ratios are available and will not be summarised here.


\begin{figure}[t] \centering \includegraphics[width=\textwidth]{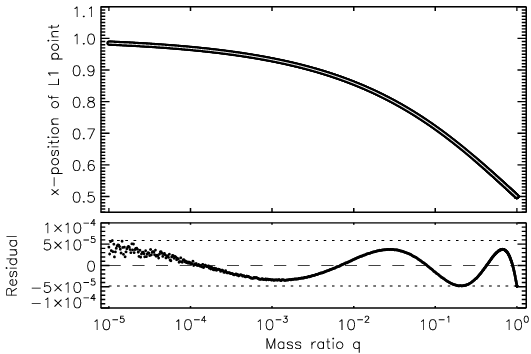} \\
\caption{\label{fig:l1:me} Top: $x$-position of the L1 point (black filled circles)
compared to the approximation in Eq.~\ref{eq:mine:l1} (white line) on a logarithmic
scale. Bottom: residuals of the approximation to the true values, with the values of
the largest positive and negative residuals indicated with dotted lines.} \end{figure}

\section*{New approximations to the positions of the L1, L2 and L3 points}

We were motivated to obtain analytic approximations to the positions of the L1, L2 and L3 points with a higher precision that existing approximations. This required a grid of positions as a function of mass ratio and a fit to this grid of various analytical functions.

After some experimentation we set up a grid of positions as a function of mass ratio containing two components. The first component consisted of 448 points which were spaced equally in $\log q$ in the interval $q = [10^{-5},0.295]$. The second component consisted of 141 points spaced equally in $q$ in the interval $q = [0.3,1.0]$. The final point at $q=1$ was given ten times the weight of the other points, as this was found to decrease the amount of `flailing' at this extremum. The positions of the Lagrangian points at each $q$ were obtained to high precision using grid search methods. We then experimented with a range of possible analytical formulae including numerical coefficients which were optimised using the {\sc mpfit} package from Craig Markwardt \cite{Markwardt07aspc}.

After some experimentation, we found that a good approximation for the position of L1 is
\begin{equation} \label{eq:mine:l1}
x_{\rm L1} = 1 - \frac{q^{a_1}}{a_2q^{a_3} + a_4}
\end{equation}
where $a_i$ are the numerical coefficients to optimise. The values of the coefficients are $a_1 = 0.33071$, $a_2 = 0.51233$, $a_3 = 0.49128$ and $a_4 = 1.487864$, and the largest deviation from the true position is $5.9 \times 10^{-5}$. For convenience, we have attempted to limit the number of significant figures required for these coefficients. Our choice of this function was inspired by that for the Roche lobe radius by Eggleton \cite{Eggleton83apj}. The fit and residuals are shown in Fig.~\ref{fig:l1:me}.

\begin{figure}[t] \centering \includegraphics[width=\textwidth]{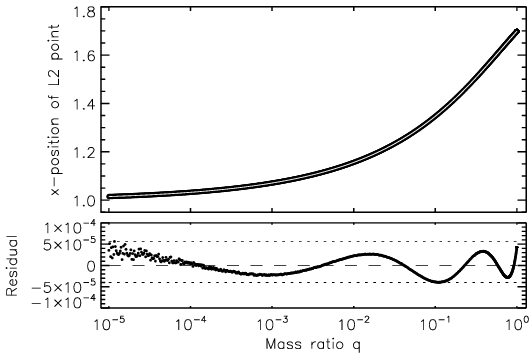} \\
\caption{\label{fig:l2:me} As for Fig.~\ref{fig:l1:me} but for the L2
approximation in Eq.~\ref{eq:mine:l2}.} \end{figure}

Approximating the position of L2 required a more complex formula. We found
\begin{equation} \label{eq:mine:l2}
x_{\rm L2} = 1 + \frac{q^{b_1} + b_2q^{b_3}}{b_4q^{b_5} + b_6}
\end{equation}
where the coefficients are $b_1 = 0.8383$, $b_2 = 2.891$, $b_3 = 0.3358$, $b_4 = 1.525$, $b_5 = 0.848$ and $b_6 = 4.046596$. The largest deviation in this case is $5.6 \times 10^{-5}$ (Fig.~\ref{fig:l2:me}). We have again limited the number of decimal places required for all coefficients except $b_6$; a slightly more precise result (maximum deviation $5.5 \times 10^{-5}$) could be obtained without this restriction.

\begin{figure}[t] \centering \includegraphics[width=\textwidth]{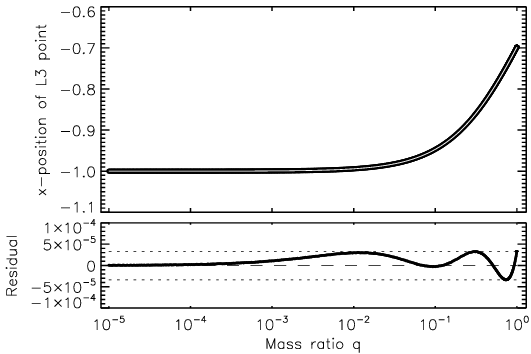} \\
\caption{\label{fig:l3:me} As for Fig.~\ref{fig:l1:me} but for the L3
approximation in Eq.~\ref{eq:mine:l3}.} \end{figure}

For L3 we found the formula
\begin{equation} \label{eq:mine:l3}
x_{\rm L3} = -1 + \frac{q^{c_1}}{c_2q^{c_3} + c_4}
\end{equation}
to be adequate, where $c_1 = 1.007$, $c_2 = 1.653$, $c_3 = 0.9375$ and $c_4 = 1.66308$. The largest deviation is $3.3 \times 10^{-5}$ (see Fig.~\ref{fig:l3:me}) but this could be lowered to $3.2 \times 10^{-5}$ if the number of decimal places is not restricted for $c_1$, $c_2$ and $c_3$.


\section*{Final comments}

The positions of the five Lagrangian points are of interest in many areas of celestial mechanics, in particular the positions of space missions such as \textit{Gaia} and JWST. The L1, L2 and L3 points can be found by solving a fifth-order polynomial, which has no analytic solution so must be performed numerically. There are times when an approximate answer is adequate, but existing analytical approximations are designed for small mass ratios.

We have therefore derived new approximations for the positions of the L1, L2 and L3 points, which can be used for convenience when an exact answer is not needed. They are accurate to within $6 \times 10^{-5}$ (absolute deviations) for all mass ratios from $10^{-5}$ to unity. Although we have taken care to limit the required precision of the numerical coefficients to four decimal places where possible, some must be specified to six decimal places to achieve the quoted precision.

Our new approximations for the L1 and L2 positions perform less well at the smallest mass ratios. For mass ratios less than $10^{-5}$ we recommend that the Hill sphere approximations are used instead.


\section*{Acknowledgements}

We are grateful to Dr.\ Andrej Pr\v{s}a for useful discussions, and to the referees for their comments and help. The NASA Astrophysics Data System was used in the course of this work.



\begin{thebibliography}{10}
\newcommand{\enquote}[1]{`(#1)'}

\bibitem{Kopal59book}
Z.~{Kopal}, \textit{{Close Binary Systems}} (The International Astrophysics
  Series, London: Chapman \& Hall), 1959.

\bibitem{Bate++71book}
R.~R. {Bate}, D.~D. {Mueller} \& J.~E. {White}, \textit{{Fundamentals of
  astrodynamics}} (Dover Publications Inc., New York, US), 1971.

\bibitem{Hilditch01book}
R.~W. {Hilditch}, \textit{{An Introduction to Close Binary Stars}} (Cambridge
  University Press, Cambridge, UK), 2001.

\bibitem{Roy05book}
A.~E. {Roy}, \textit{{Orbital motion, 4th edition}} (Institute of Physics
  Publishing, Bristol, UK), 2005.

\bibitem{Prsa18book}
A.~{Pr{\v{s}}a}, \textit{{Modeling and Analysis of Eclipsing Binary Stars; The
  theory and design principles of PHOEBE}} (IoP Publishing, Bristol, UK), 2018.

\bibitem{LeahyLeahy15comac}
D.~A. {Leahy} \& J.~C. {Leahy}, \textit{Computational Astrophysics and
  Cosmology}, \textbf{2}, 4, 2015.

\bibitem{DepaterLissauer01book}
I.~{de Pater} \& J.~J. {Lissauer}, \textit{Planetary Sciences, by Imke de Pater
  and Jack J.~Lissauer, Cambridge University Press, Cambridge, UK}, 2001.

\bibitem{Battin87book}
R.~H. {Battin}, \textit{{An Introduction to the Mathematics and Methods of
  Astrodynamics}} (American Institute of Aeronautics and Astronautics,
  Virginia, US), 1987.

\bibitem{MurrayDermott00book}
C.~D. {Murray} \& S.~F. {Dermott}, \textit{{Solar system dynamics}} (Cambridge
  University Press, Cambridge, UK), 2000.

\bibitem{Markwardt07aspc}
C.~B. {Markwardt}, 2009, \textit{Astronomical Society of the Pacific Conference
  Series}, vol. 411, p. 251.

\bibitem{Eggleton83apj}
P.~P. {Eggleton}, \textit{ApJ}, \textbf{268}, 368, 1983.

\end{thebibliography}

\end{document}